\documentclass[12pt,preprint]{aastex}
\usepackage{natbib}
\def\snr{SN\,1993J~}
\def\sn1006{SN\,1006~}
\def\sigmad{$\Sigma-D~$}
\def\aap{A\&A\,  }
\def\aj{AJ  }
\def\apj{ApJ\,  }
\def\apjl{ApJ\,  }
\def\apjs{ApJS  }
\def\apss{Astrophysics and Space Science  }
\def\araa{ARA\&A  }
\def\azh{AZh}
\def\cjaa{Chinese J. Astron. Astrophys.  }

\def\mnras{MNRAS\,  }

\def\rmp{Rev. Mod. Phys.  }

\begin{document}
\shorttitle
{
Surface-brightness diameter in SNR
}
\shortauthors{Zaninetti}
   \title{
Analytical and Monte Carlo results
for the surface-brightness diameter
relationship  in supernova remnants
          }
\author{Lorenzo Zaninetti  }

\affil {Dipartimento di Fisica Generale, Via Pietro Giuria 1,\\
           10125 Torino, Italy}

\email   {zaninetti@ph.unito.it  \\
\url     {http://www.ph.unito.it/$\tilde{~}$zaninett}}
\begin{abstract}
The  surface  brightness  diameter  relationship for  supernovae
remnants  (SNRs) is  explained by adopting a model  of direct
conversion  of the flux of kinetic energy  into synchrotron
luminosity. Two laws of motion are  adopted, a power law model
for the radius-time relationship, and a model  which uses the
thin layer approximation. 
The  fluctuations on the log-log
surface diameter relationship  are modeled by a  Monte Carlo
simulation.
In this  model  a 
new probability density function for the
density as  function of the  galactic 
height is introduced.
\end {abstract}

\section{Introduction}

The correlation between  radio surface  brightness,
$\Sigma_{1GHz}$, and  diameter $D$ in pc
in supernova remnants  is  parametrized as
\begin{equation}
\Sigma_{1GHz}  = C_{\Sigma} (D)^{-\beta_{\Sigma}}
\quad \frac{W} {m^2 \,Hz\, sr}  \quad ,
\label{sigmaddef}
\end{equation}
where  ${\beta_{\Sigma}}$  and
$C_{\Sigma}$  are are found through
the radio observations
\citep{Shklovskii1960a,Urosevic2005}.
This  is an observational  relationship
which has  been  explained by early
theoretical models
\citep{Lequeux1962,Poveda1968,Kesteven1968,Clark1976,Milne1979}.
A theoretical interpretation of the
\sigmad relationship has  been introduced by
\cite{Duric1986}
by incorporating the Sedov
blast wave solution for supernovae  remnant (SNR)  expansion,
the generation and evolution of the
magnetic field as outlined by \cite{Gull1973},
and the acceleration of relativistic
electrons by shocks, as formulated by \cite{Bell_I,Bell_II}.
The time-dependent,  nonlinear kinetic theory
for cosmic ray (CR) production was
adopted  by  \cite{Berezhko2004}
in order  to  derive a
theoretically predicted brightness-diameter
relation in the radio range for  the Sedov phase.
Recently \cite{Bandiera2010}
found
that SNRs cease to emit effectively in radio at a stage
near the end of their Sedov evolution, and
that models of synchrotron
emission with constant efficiencies in
particle acceleration and magnetic
field amplification do not provide a close match to the data.
These previous models leave  a series of questions
unanswered or merely partially answered:
\begin{itemize}
\item Is it possible to deduce  a theoretical \sigmad relationship
on the basis of power law expansion for SNR? \item Is it possible
to deduce  a classical equation of expansion for the SNR with two
adjustable parameters  that can be found from  a numerical
analysis of the radius--time relationship? \item Is the
theoretical \sigmad relationship  sensitive to the density of the
interstellar medium, which decreases as a function of height above
the galactic plane?

\end{itemize}
In order to answer these questions, we develop a theoretical
derivation of the \sigmad relationship by modeling the expansion
of a synchrotron emitting shell. In Section \ref{sec_power}, we
show that it is reasonable to fit a power-law expansion
relationship to a 
known young supernova remnant, SN 1987J in M81, for
 which  the growth of
the shell like structure is  available over a period  of 10 years.
In Section \ref{sec_classical}, we discuss a modification of the
power-law expansion model for the effects of the standard gradient
in density in the interstellar medium. In Section
\ref{sigmadsection}, we complete our derivation of the \sigmad
relationship by adopting a direct conversion of the flux  of
kinetic energy into radiation  calibrated on the observational
data (Section \ref{sec_observations}). In Section \ref{sec_sigmadconf}, we
address the observed fluctuations in the \sigmad relationship by
utilizing a new probability density function (PDF) for the
contrast of in situ densities of SNRs. This  new PDF is derived
from the standard number distribution of SNRs with the galactic
height.

\section{The Power Law Model of Expansion}
\label{sec_power}

\begin{figure*}
\begin{center}
\plotone{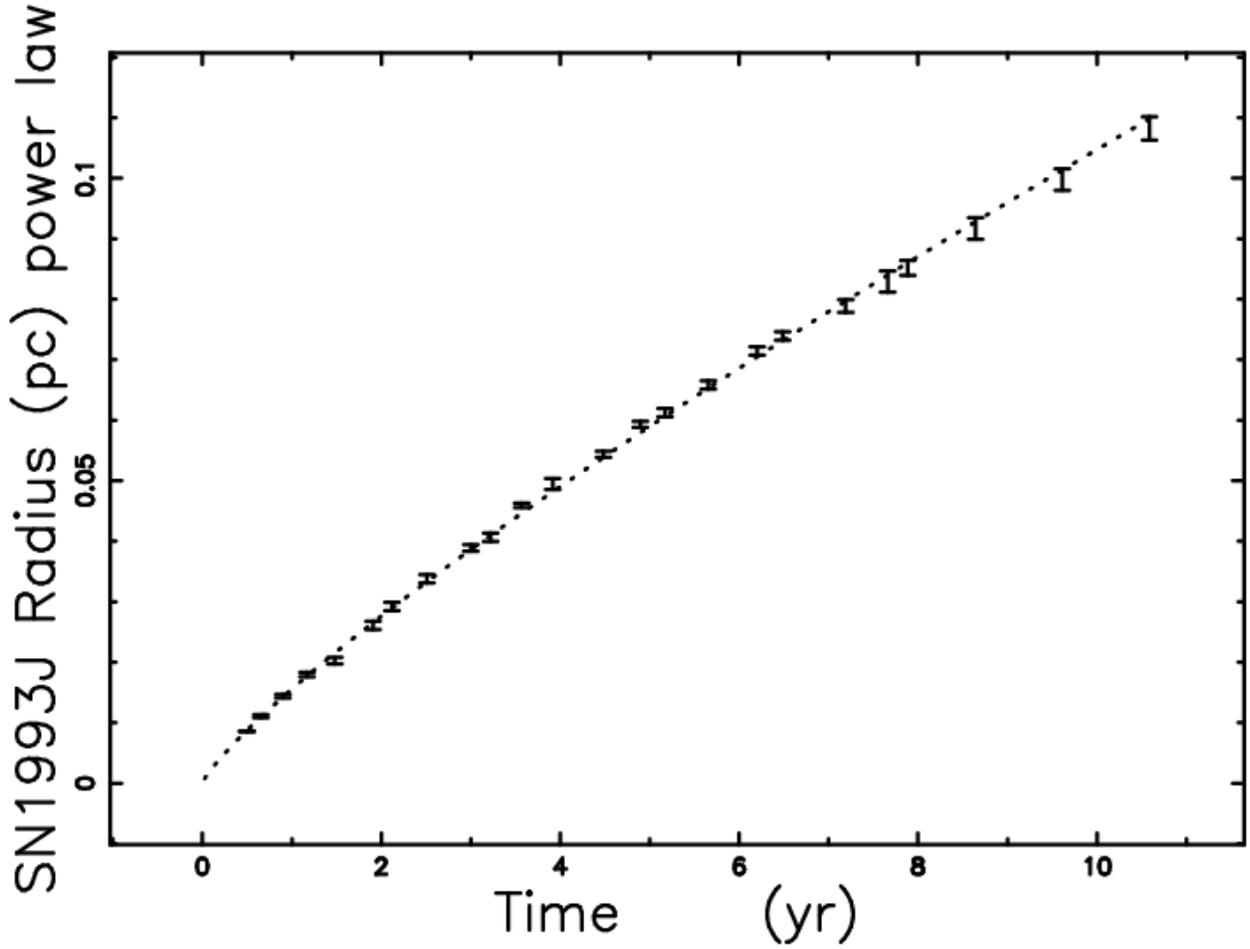}
\end {center}
\caption
{
Theoretical radius as given by the power law model
with data as in  Table~\ref{datafit} (full line),
and astronomical data of \snr with
vertical error bars.
}
\label{radiust}
    \end{figure*}

\begin{figure*}
\begin{center}
\plotone{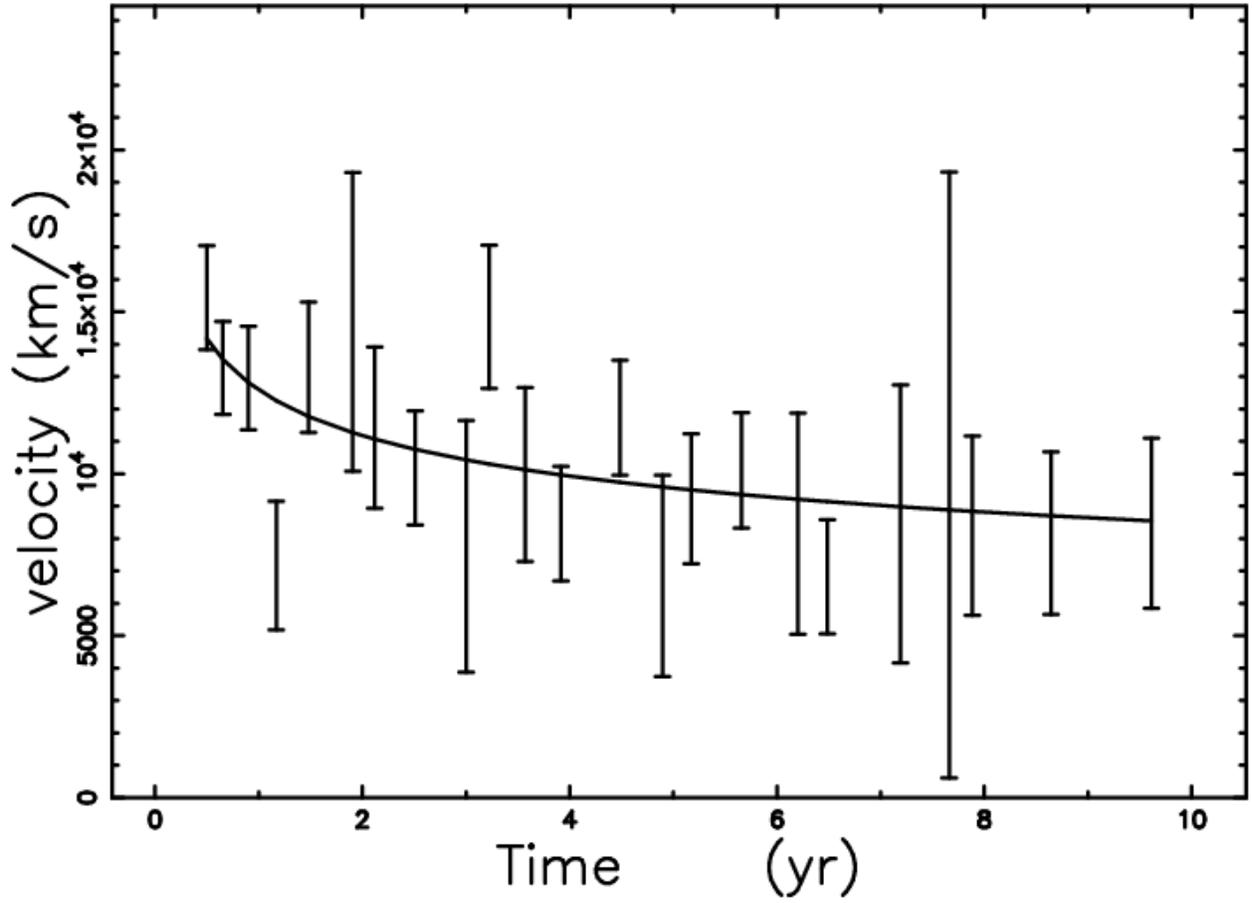}
\end {center}
\caption
{
Theoretical velocity
as  given  by Eq. (\ref{velpower})
(full  line)
and instantaneous velocity  of \snr  with
uncertainty.
}
\label{velt}
    \end{figure*}

A standard point of view  assumes that the SNR  expands at a
constant velocity until the surrounding mass is of the order of
the solar mass. 
This time scale can be modeled by
\begin {equation}
t_M= 186.45\,{\frac {\sqrt [3]{{\it M_{\sun} }}}{\sqrt [3]{{\it
n_0}}{\it v_{10000}}}} \quad yr \quad ,
\end{equation}
where $ M_{\sun}$ is the number of solar masses in the volume
occupied by the SNR, $n_0$,  the number density  expressed  in
particles~$\mathrm{cm}^{-3}$, and $v_{10000}$ the initial velocity
expressed in units of $10000\,km/s$ , see \cite{McCray1987}.
To model the time-dependent expansion of SNRs, we use
\begin{equation}
R(t) = C  t^{\alpha} \label{rpower} \quad ,
\end{equation}
where the two parameters $C$ and  $\alpha$ 
can be found from the
observations. Ten years of observations  of \snr 
\citep{Marcaide2009}  allow us  to fit these two parameters which are
reported in Table~\ref{datafit} and in Figure \ref{radiust}. The
quality of the fit is measured by the merit function
\begin{equation}
\chi^2  =
\sum_j \frac {(R_{th} -R_{obs})^2}
             {\sigma_{obs}^2}
\quad ,
\label{chisquare}
\end{equation}
where  $R_{th}$, $R_{obs}$ and $\sigma_{obs}$
are the theoretical radius, the observed radius and
the observed uncertainty respectively.
This observed relationship allows us to express the radius,
as a function of time, by
\begin{equation}
R(t) =   0.0155\,{t}^{ 0.828} \quad pc \quad,
\label{radiustime}
\end{equation}
where the  time  $t$  is  expressed in $yr$.
The  velocity  in this model  is
\begin{equation}
V(t) = 12587.67\,{t}^{- 0.171} \frac{km}{s}  \quad .
\label{velpower}
\end{equation}
Figure \ref{velt} displays the observed instantaneous velocity as
deduced from the finite difference method and the best fit
according to Eq.  (\ref{velpower}). Another   interesting  SNR  is
\sn1006 which started to  be  visible  in 1006 AD and has  an
actual radius of  $15^{\prime}$ according to \cite{Green2009}. 
The
radius in pc is a function  of the the adopted distance; as 
an   example,
\cite{Katsuda2010} quotes a distance of  2.2 kpc  
and \cite{Winkler2003} 2.18
kpc. 
Here  we adopt a distance  of 2.2\,kpc  and consequently
the  observed radius, $R_{obs}$, is
\begin{equation}
R = 9.59 D_{22}\, pc
\quad ,
\end{equation}
where  the  $D_{22}$  is  the distance
expressed in
units  of  2.2 kpc.
The proper motion of the remnant's edge is a weak
  function
of the azimuth angle and the average value is
\begin{equation}
\overline {P_{obs}} =
0.5\,  arcsec /yr
\quad .
\end{equation}
The average  velocity of the expansion is
\begin{equation}
V_{obs} =  5441 \,D_{22} \frac{km}{s}  \quad .
\end{equation}
Eq. (\ref{rpower}) and  the connected velocity can  be  solved for
the two unknown  variables $\alpha$ and  $C$
\begin{eqnarray}
\alpha =
 0.000001022\,{\frac {{\it V_{obs}}\,t}{{\it R_{obs}}}}
\nonumber  \\
C =
\frac {\it R_{obs}}
{{t}^{ 0.000001022\,{\frac {{\it V_{obs}}\,t}{{\it R_{obs}}}}}}
\label{twovar}
\end{eqnarray}
where  $V_{obs}$  is  expressed in $\frac{km}{s}$ , $R_{obs}$ in
pc and $t$ in yr. Table~\ref{datavar}  reports  the numerical
results. 

\section{Momentum conservation and  galactic height}

\label{sec_classical}

The presence of a slow wind from the SN progenitor 
  makes it reasonable for us to assume
 that the SNR evolves in a
previously ejected medium phase in which density is considerably
higher  than the interstellar medium (ISM). The assumption 
used here is that the density of the ISM around the SNR has the
following two piecewise dependencies
\begin{equation}
 \rho (R)  = \left\{ \begin{array}{ll}
            \rho_0                      & \mbox {if $R \leq R_0 $ } \\
            f\,\rho_0 (\frac{R_0}{R})^d    & \mbox {if $R >    R_0 $ ~.}
            \end{array}
            \right.
\label{piecewise}
\end{equation}
where   $f$  is  a parameter that  models the  jump in density ,
$0 < f \leq 1$, and is connected with the  vertical profile  in
density as a  function of the  galactic height. In this framework,
the density decreases as an inverse power law with an exponent $d$
that can be fixed from the observed temporal evolution of the
radius, with $d=0$ meaning constant density;
 further on the
parameter $f$ regulates  the density at  $R=R_0$.
The swept mass
in the interval $0 \leq r  \leq R_0$
is
\begin{equation}
M_0 =
\frac{4}{3}\,\rho_{{0}}\pi \,{R_{{0}}}^{3}
\quad .
\end{equation}
The
swept
mass
in the interval $0 \leq r \leq R$
with $r \ge R_0$
is
\begin{eqnarray}
M =
-4\,f\,{r}^{3}\rho_{{0}}\pi \, \left( {\frac {R_{{0}}}{r}} \right) ^{d}
 \left(d -3 \right) ^{-1}   \nonumber \\
+4\,f\,{\frac {\rho_{{0}}\pi \,{R_{{0}}}^{3}}{d-
3}}
+ \frac{4}{3}\,\rho_{{0}}\pi \,{R_{{0}}}^{3}
\quad .
\end{eqnarray}
Momentum conservation requires  that
\begin{equation}
M v = M_0 v_0
\quad ,
\end {equation}
where  $v$   is  the velocity at $t$
and    $v_0$ is  the velocity at $t=t_0$.
The velocity  as a function of the radius
is
\begin{equation}
v  =
\frac{
v_{{0}}{r_{{0}}}^{3} \left( 3-d \right)
}
{
3\,{r_{{0}}}^{d}{R}^{3-d}f+{r_{{0}}}^{3} \left( -d+3-3\,f \right)
}
\quad .
\label{velclassic}
\end{equation}
In this first order differential equation in R,
the
variables can be separated and an integration
term by term yields
the following
nonlinear equation
\begin{eqnarray}
{\mathcal{F}}_{NL} = \nonumber \\
 \left( 12\,{R_{{0}}}^{3}f-3\,{R_{{0}}}^{3}fd+7\,{R_{{0}}}^{3}d
-{R_{{0
}}}^{3}{d}^{2}-12\,{R_{{0}}}^{3} \right) R
\nonumber \\
-3\,{R_{{0}}}^{d}f{R}^{4-d}+
7\,{R_{{0}}}^{3}v_{{0}}dt_{{0}}+{R_{{0}}}^{3}v_{{0}}{d}^{2}t
\nonumber \\
-{R_{{0}}}
^{3}v_{{0}}{d}^{2}t_{{0}}+{R_{{0}}}^{4}{d}^{2}+12\,{R_{{0}}}^{4}
-9\,{R
_{{0}}}^{4}f-7\,{R_{{0}}}^{4}d-7\,{R_{{0}}}^{3}v_{{0}}dt
\nonumber  \\
-12\,{R_{{0}}}
^{3}v_{{0}}t_{{0}}+3\,{R_{{0}}}^{4}fd+12\,{R_{{0}}}^{3}v_{{0}}t
\quad  .
\label{nonlinearf}
\end {eqnarray}
An approximate solution of
${\mathcal{F}}_{NL}(r) $  can be obtained
assuming that   \newline
$3 R_0^d R^{4-d} f $
$\gg$
$ {R_{{0}}}^{3} \left( 4-d \right)  \left( d-3+3\,f \right) R $
\begin{eqnarray}
 R(t) = \nonumber \\
 ( {R_{{0}}}^{4-d}-\frac{1}{3f}(d-3+3f){R_{{0}}}^{4-d}
( 4-d ) \nonumber \\
 + \frac{1}{3f}
 ( 4-d ) v_{{0}}{R_{{0}}}^{3-d} ( 3-d )
 ( t-t_{{0}} )  ) ^{\frac{1}{4-d}}
\quad .
\label{asymptotic}
\end{eqnarray}

The physical units have not been specified, $pc$ for length and
$yr$ for time are perhaps an acceptable astrophysical choice. With
these units, the initial velocity $v_{{0}}$ is expressed in
$\frac{pc}{yr}$ and should be converted into $\frac{km}{s}$; this
means that $v_{{0}} =1.02\,10^{-6} v_{{1}}$ where  $v_{{1}}$ is
the initial velocity expressed in $\frac{km}{s}$.

\begin{figure*}
\begin{center}
\plotone{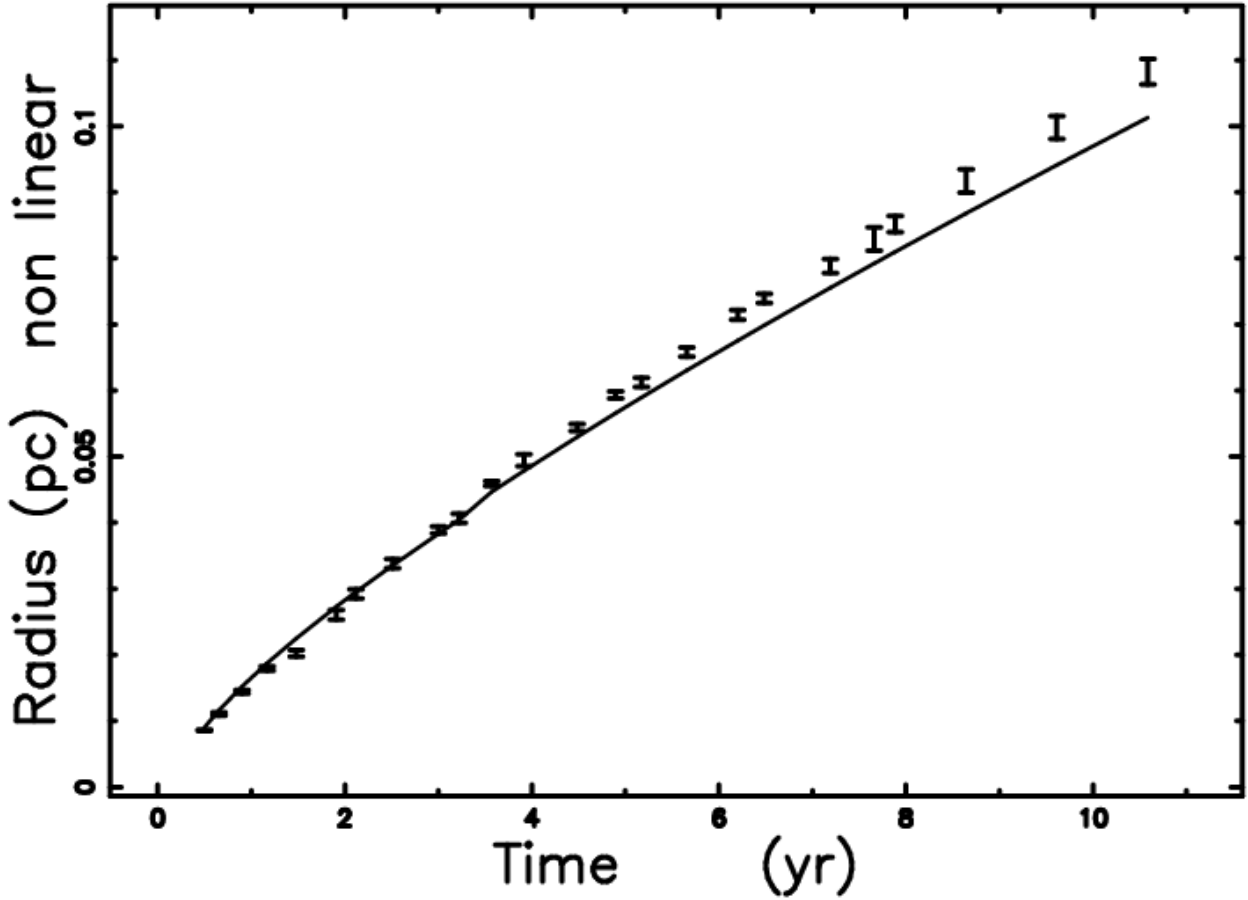}
\end {center}
\caption
{
Theoretical radius as obtained
by the solution of  the nonlinear
Eq. (\ref{nonlinearf})
(full line), data as in Table~\ref{datafit} and  $f$=1.
The astronomical data of \snr are represented with
vertical error bars.
}
\label{1993pc_fit_nl}
    \end{figure*}

\begin{figure*}
\begin{center}
\plotone{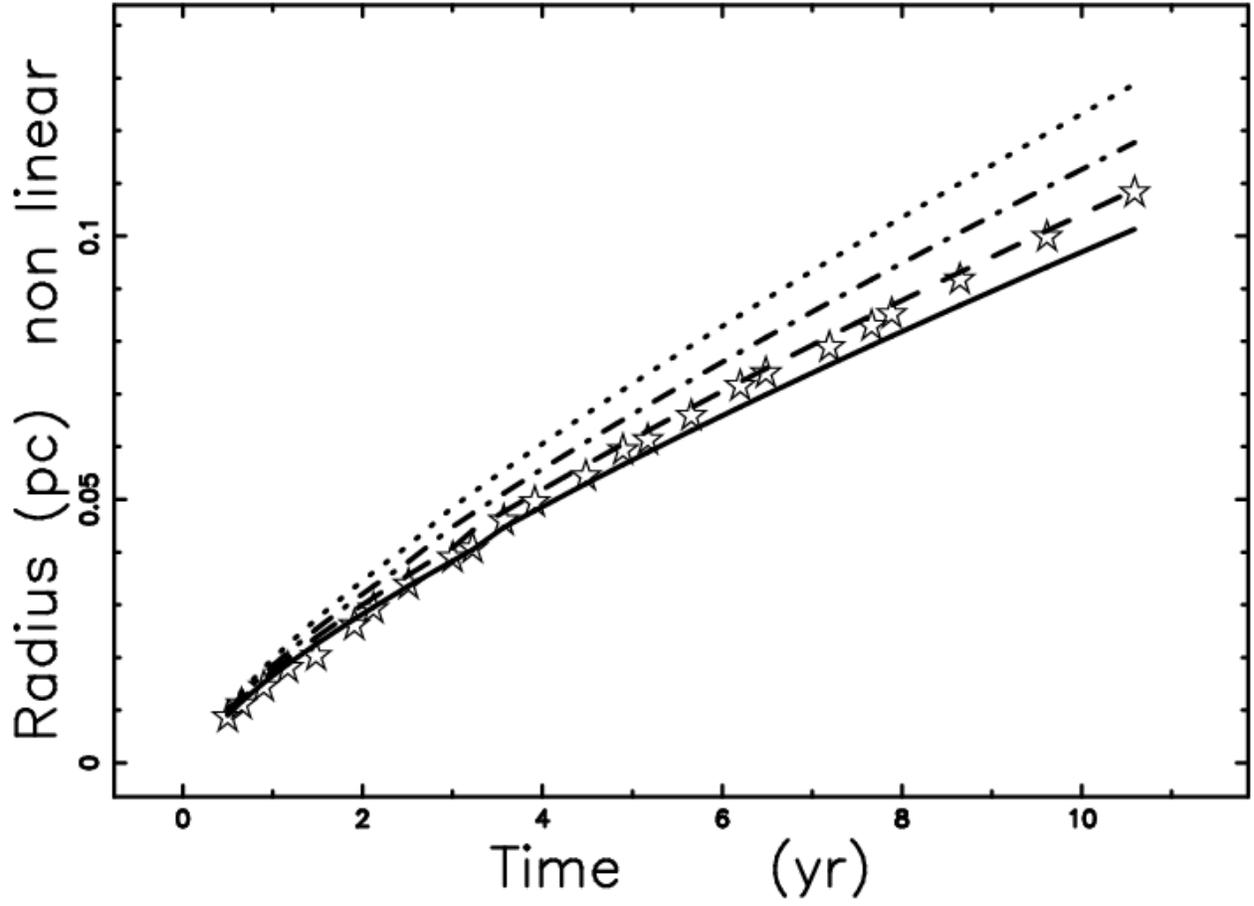}
\end {center}
\caption
{
The same as Figure \ref{1993pc_fit_nl}
but
$f$ is  variable;
$f=1$    (full line),
$f=0.9$  (dashed)   ,
$f=0.8$  (dot-dash-dot-dash)
$f=0.7$  (dotted)
and  experimental point represented by empty stars.
}
\label{1993pc_fit_nlf}
    \end{figure*}

A one-dimensional solution of  Eq. (\ref{nonlinearf}) 
can be found 
with the FORTRAN subroutine ZRIDDR \citep{press}.  
A plot of this radial solution, using a parameter value 
of f = 1, is shown in Figure (\ref{1993pc_fit_nl}).  
Plots of other solutions with different value of f 
can be found in Figure (\ref{1993pc_fit_nlf}).


\section{\sigmad in the power law model}
\label{sigmadsection}
In this section, we review the time scale of 
synchrotron losses, 
the time scale of acceleration and the 
inequality which allows us to use
the in situ acceleration of electrons. The relation
between the radio surface  brightness and diameter of SNRs  is
deduced assuming a direct  proportionality   between  the
radio-luminosity and the flux of kinetic energy.

\subsection{The  synchrotron emission}

An electron which  loses
its  energy  due to
synchrotron radiation
has a lifetime  of
\begin{equation}
\tau_r  \approx  \frac{E}{P_r} \approx  500  E^{-1} H^{-2} sec
\quad ,
\label {taur}
\end{equation}
where
$E$  is the energy in ergs,
$H$ the magnetic field in Gauss,
and
$P_r$  is the total radiated
power \citep[] [Eq. 1.157] {lang}.
The  energy  is connected  to  the critical
frequency \citep[][Eq. 1.154]{lang},
as
\begin {equation}
\nu_c = 6.266 \times 10^{18} H E^2~Hz
\quad  .
\label {nucritical}
\end{equation}
The lifetime
for synchrotron  losses is
\begin{equation}
\tau_{syn} =
 39660\,{\frac {1}{H\sqrt {H\nu}}} \, yr
\quad  .
\end{equation}

Following \citep{Fermi49,Fermi54},
the gain  in  energy  in a continuous    form
for  a particle
which  spirals  around a line of force
is  proportional to its
energy, $E$,
\begin  {equation}
\frac {d  E}  {dt }
=
\frac {E }  {\tau_{II} }   \quad,
\end {equation}
where $\tau_{II}$ is the  typical time-scale,
\begin {equation}
\frac{1}{\tau_{II}}  = \frac {4} {3 }
( \frac {u^2} {c^2 }) (\frac {c } {L })
\quad ,
\label {tau2}
\end   {equation}
where $u$ is the velocity of the accelerating cloud,
$c$  is the velocity of the light
and $L$  is the
mean free path between clouds \citep[][Eq. 4.439]{lang}.
The mean free path between the accelerating clouds
in the Fermi~II mechanism can be found from the following
inequality in time :
\begin{equation}
\tau_{II} < \tau_{sync}
\quad  ,
\end{equation}
which  corresponds to  the following  inequality for the
mean free  path  between scatterers
\begin{equation}
L  <
\frac
{
1.72\,10^5 \,{u}^{2}
}
{
H\sqrt {H\nu}{c}^{2}
}
\,pc
\quad  .
\label{fundamental}
\end{equation}
The mean free path length for \snr
at  $\nu=1GHz$
gives
\begin{equation}
L <
1.83\,10^{-5}{t}^{- 0.34}
\, pc
\end{equation}
where  the velocity  is given  by Eq.(\ref{velpower})
and  $H =65.1 \,G  $
\citep{MartiVidalb2011}.
When this inequality  is verified  the direct
conversion of the flux  of  kinetic energy
into radiation can be adopted.
Recall that the Fermi~II  mechanism  produces
an  inverse power law
spectrum in the
energy
of the type
$
N (E) \propto  E ^{-\gamma}
$
or an  inverse power law  in the observed frequencies
$
N (\nu) \propto   \nu^{-\alpha }
$
with  $ \alpha= \frac{\gamma -1}{2}$
\citep{lang,Zaninetti2011a}.

\subsection{The dimensional approach}

\label{sec_sigmad}
The source of synchrotron luminosity
is here assumed to be
the flux of kinetic energy
\begin{equation}
L_m = \frac{1}{2}\rho A  V^3
\quad ,
\end{equation}
where A is the emitting surface area 
\citep[][Eq. A28]{deyoung}.
Assuming the surface area is like that of a sphere, the luminosity
is
\begin{equation}
L_m = \frac{1}{2}\rho 4\pi R^2 V^3
\quad ,
\end{equation}
where $R$  is the instantaneous radius of the SNR and
$\rho$  is the density in the advancing layer
in which the synchrotron emission takes place.
Assuming the density of the advancing layer scales as
$R^{-d}$,
which means that
\begin{equation}
L_m  \propto R ^{2-d}  V^3
\quad .
\end{equation}
The time dependence is eliminated by utilizing Eq.  (\ref{rpower})
\begin{equation}
L_m = L_0 (\frac{R}{R_0})^{-{\frac {d\alpha-5\,\alpha+3}{\alpha}}}
\quad ,
\label{lumtheo}
\end{equation}
where  $L$  is  the luminosity at $R=R_0$
or
\begin{equation}
L_m = L_0 (\frac{D}{D_0})^{-{\frac {d\alpha-5\,\alpha+3}{\alpha}}}
\quad ,
\end{equation}
where $D$  is the actual  diameter and  $D_0=2R_0$.
On assuming  that  the observed luminosity ,$L_{\nu}$,
 in a given band
denoted by the frequency  $\nu$ is  proportional
to the mechanical  luminosity
we obtain
\begin{equation}
L_{\nu}  = L_{0,\nu} (\frac{D}{D_0})^{-{\frac {d\alpha-5\,\alpha+3}{\alpha}}}
\quad Jy\, kpc^2
\quad ,
\end{equation}
where  $L_{0,\nu}$ is the observed
radio luminosity in a given band.
The observations are  generally   represented
in the form
\begin{equation}
L = C_L D^{-\beta_L}
\quad ,
\end{equation}
where  $L$ and  $C_L$ are parameters deduced
from the radio observations \citep{Guseinov2003,Urosevic2005}.
Given the two observational parameters $\alpha$ and $\beta_L$
we can derive
\begin{equation}
d = {\frac {\beta_L\,\alpha+5\,\alpha-3}{\alpha}}
\quad .
\end{equation}

The radio surface  brightness  of a remnant
is  defined as
\begin{equation}
\Sigma =  \frac{S_{1GHz}} {\theta^2}
\quad  ,
\end{equation}
where  $S_{1GHz}$   is the detected  flux  of a remnant
at $1GHz$ and  $\theta$  the observed angle.
Due to the fact  that
$\theta \propto D^2$ we  have
\begin{equation}
\Sigma_{1GHz}  = \Sigma_{0,1GHz}
(\frac{D}{D_0})^{-{\frac {d\alpha-3\,\alpha+3}{\alpha}}}
\quad \frac{W} {m^2 \,Hz\, sr} \quad ,
\label{surftheo}
\end{equation}
where  $\Sigma_{0,{1GHz}}$ is the surface brightness
at  $D=D_0$ .
According to the basic  observational
relationship (\ref{sigmaddef})
\begin{equation}
d=
{\frac {{\it \beta_{\Sigma}}\,\alpha+3\,\alpha-3}{\alpha}}
\quad .
\end{equation}

\section{Observations in the  power law model}

\label{sec_observations}

\begin{figure*}
\begin{center}
\plotone{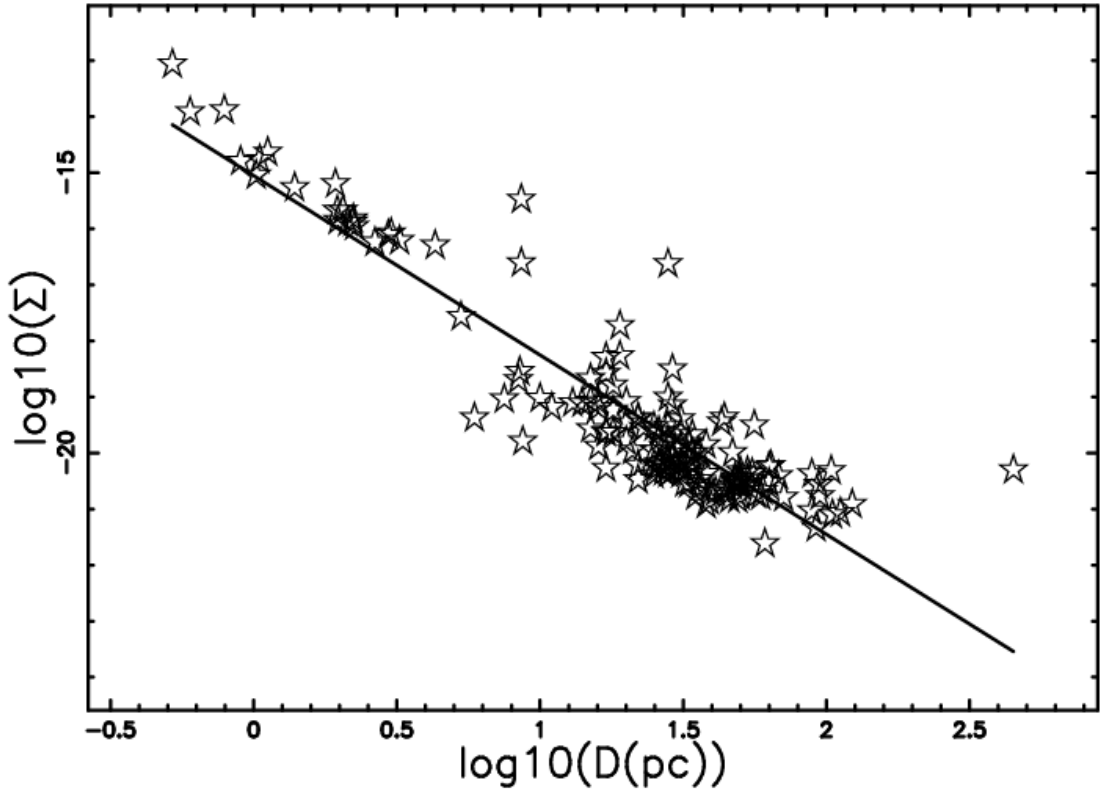}
\end {center}
\caption
{
The  LOG-LOG \sigmad diagram at a frequency of
1 GHz for all the extragalactic  SNRs  as given
by  \cite{Urosevic2005}.
}
          \label{sigmadextra}%
    \end{figure*}

Supernova remnants in our galaxy have
two luminosity-diameter relationships  \citep{Guseinov2003}:
the first one for SNRs which have
L $>$
5300 Jy kpc$^2$, D $<$ 36.5 pc and the second
one  for  SNRs having L $\le$
5300 Jy kpc$^2$, D $\ge$ 36.5 pc,
\begin{equation}
L=2.45 \, 10^4 D^{-0.43}
\hspace{0.3cm} Jy \hspace{0.1cm} kpc^2
\quad  D  <  36.5 pc  \quad  ,
\end{equation}
and
\begin{equation}
L=5.38 \, 10^9 D^{-3.84} \hspace{0.3cm} Jy \hspace{0.1cm} kpc^2
\quad  D  \ge  36.5 pc  \quad .
\end{equation}
The  \sigmad  relationship  in a similar way is
\begin{equation}
\Sigma=2.7^{+2.1}_{-1.4} \, 10^{-17} D^{{-2.47}^{+0.20}_{-0.16}}
\hspace{0.3cm} W m^{-2} Hz^{-1} ster^{-1}
\quad  D  <  36.5 pc  \quad  ,
\end{equation}
and
\begin{equation}
\Sigma=8.4^{+19.5}_{-6.3} \, 10^{-12} D^{{-5.99}^{+0.38}_{-0.33}}
\hspace{0.3cm} W m^{-2} Hz^{-1} ster^{-1}
\quad  D  \ge  36.5 pc  \quad .
\end{equation}
The  observations of the surface  brightness of   SNRs
in other galaxies has  been analyzed by \cite{Urosevic2005}
with  data available at  Centre de Données
astronomiques de Strasbourg
(CDS).
Figure \ref{sigmadextra} reports  the observational
data as well the  fitting curve  of all the radio
SNRs  in external galaxies.
The results  of our numerical
analysis
for extragalactic  SNRs
 gives
\begin{equation}
\Sigma=(8.8 \pm 3.27   \, 10^{-16} D^{-3.1 \pm 0.11}
\hspace{0.3cm} W m^{-2} Hz^{-1} ster^{-1}
\quad  D  <  450  pc  \quad   ,
\end{equation}
which agrees with the results  of  \cite{Urosevic2005}.
The values of  $d$ from  the \\
 \sigmad relationship are
reported  in  Table~\ref{datad}. 

\section{A Monte Carlo model for  \sigmad }
\label{sec_sigmadconf}

The significant   fluctuations  observed in the 
\sigmad relationship can
be explained by correlating the well known probability  to have a
SNR at the galactic height $z$ with the density. This  conversion
can be done using the  nonlinear relationship between  $z$ and
density as  given by  the self-gravitating disk.

 The radius function  we derived in Section
\ref{sec_classical}  (Eq. \ref{asymptotic}) from momentum
conservation allows us to deduce an analytical expression for the
\sigmad relationship:
\begin{equation}
\Sigma_{1GHz}  = \Sigma_{0,1GHz}
(\frac{1}{f^3}) (\frac{D}{D_0})^{-{\frac {d\alpha-3\,\alpha+3}{\alpha}}}
\quad \frac{W} {m^2 \,Hz\, sr} \quad .
\label{sigmadconf}
\end{equation}
The  \sigmad  relationship
 now   has   an inverse  cubic dependence
for   the
contrast  parameter $f$.

\subsection{The  profile  of the ISM}

\begin{figure}
  \begin{center}
\plotone{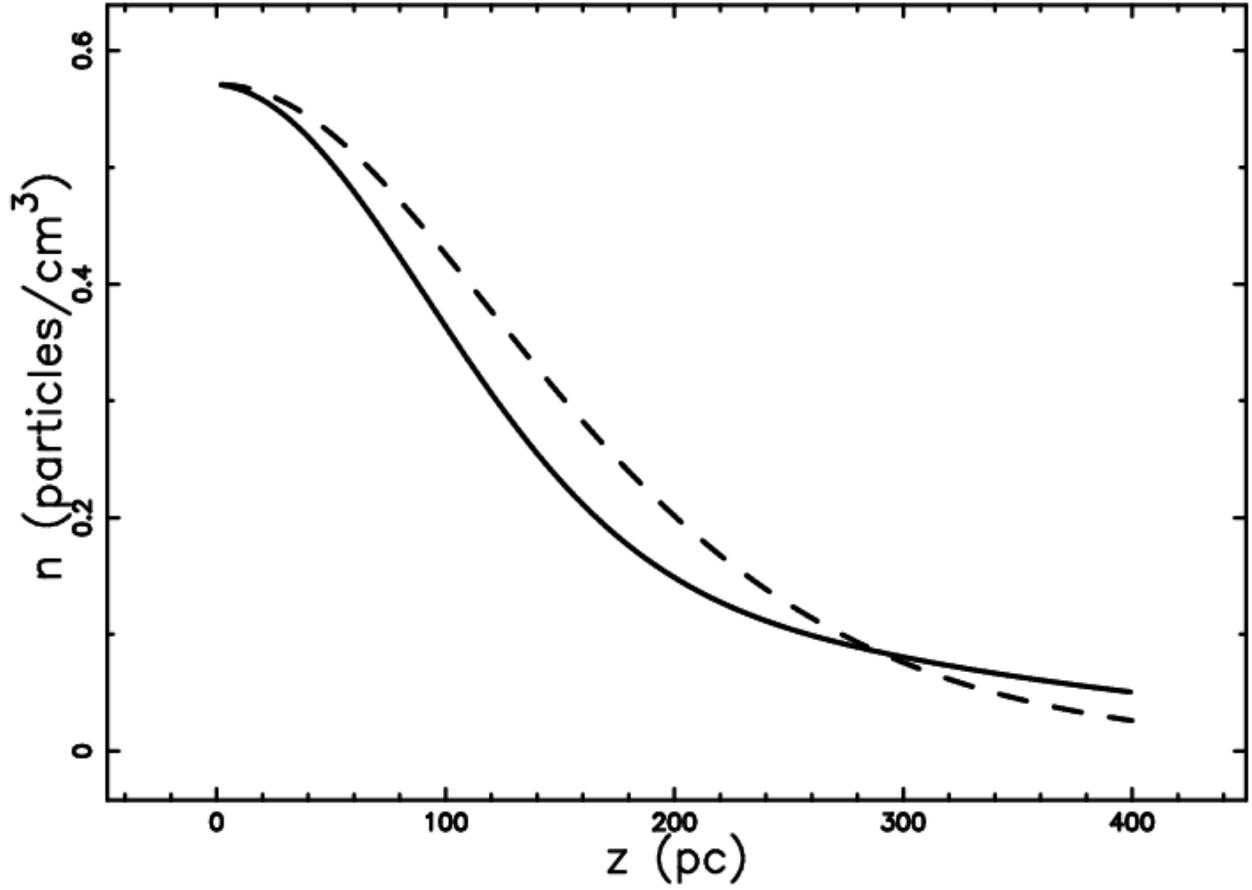}
  \end {center}
\caption
{
Profiles of density versus scale height $z$:
the  self-gravitating disk as
given by Eq. (\ref{sech2})
when $z_0=90~pc$
(dashed)
and
the
three-component exponential distribution
as
given by Eq. (\ref{exponential})
(full line).
}%
    \label{zprofile}
    \end{figure}

The vertical number density distribution
of galactic H\,I   has the following
three component  behavior as a function of
the  galactic height
{\it z} ,
the  distance  from  the galactic plane in pc
:
\begin{equation}
n(z)  =
n_1 e^{- z^2 /{H_1}^2}+
n_2 e^{- z^2 /{H_2}^2}+
n_3 e^{-  | z |  /{H_3}}
\,.
\label{exponential}
\end{equation}
We set the densities in Eq. (\ref{exponential}) to $n_1 = 0.395$
particles cm$^{-3}$, $n_2 = 0.107$ particles cm$^{-3}$, $n_3 =
0.064$ particles cm$^{-3}$, and the scale heights to $H_1 = 127$
pc, $H_2 = 318$ pc, and $H_3 = 403$
\citep{Lockman1984,Dickey1990,Bisnovatyi1995}. 
This  distribution
of  galactic H\,I is valid in the range 0.4 $\leq$  $R$ $\leq$
$R_0$, where  $R_0$ = 8.5 \mbox{kpc} and $R$  is the distance
from  the galaxy center. The  previous empirical relationship can
be modelled by  a theoretical one. The density profile of a thin
self-gravitating disk of gas which is characterized by a
Maxwellian distribution in velocity and  distribution which varies
only in the z-direction has the following number density
distribution
\begin{equation}
n(z) = n_0 sech^2 (\frac{z}{2*z_0})
\quad  ,
\label{sech2}
\end{equation}
where $n_0$ is the density at $z=0$ and
$z_0$ is a scaling parameter 
\citep{Bertin2000,Padmanabhan_III_2002}.
Figure (\ref{zprofile})  displays  a comparison between
the  empirical   function  sum of three exponential disks
and the theoretical  function
as given by the Eq.  (\ref{sech2}).
The previous  equation  can be  expressed  with our
contrast  parameter $f$
\begin{equation}
f = sech^2 (\frac{z}{2*z_0})
\quad  ,
\label{f}
\end{equation}
and  the  inversion gives   $z$ as  function of  $f$
\begin{equation}
z=2\,{\it arcsech} \left( \sqrt {f} \right) z_{{0}}
\quad  .
\label{zfunf}
\end{equation}

\subsection{The  distribution of SNRs}

The  probability   density function (PDF) ,$p(z)$,
to have a SNR as function of  the galactic height $z$
is  characterized   by an exponential   PDF
\begin{equation}
p(z)  = \frac{1}{b} \exp {-\frac{z}{b}}
\quad  ,
\label{pdfsnrz}
\end{equation}
with  $b=83$ pc \citep{Xu2005}.
The   average   value    $\overline{z}$
for SNRs
is
\begin{equation}
\overline{z} =b  = 83 pc
\quad .
\end{equation}
We briefly review   how a
PDF
$p(z)$  changes to
$g(f)$  when a new variable $f (z)$  is introduced.
The rule for
transforming  a PDF  is
\begin{equation}
g(f) =  \frac  {p (f(z)) } {
\vert\frac {d(g(f)) } {dz} \vert}
\label{trans}
\quad.
\end{equation}
Once the   previous rule is  implemented
we obtain
\begin{equation}
g(f)   =
\frac
{
{{\rm e}^{-2\,{\frac {{\it arcsech}
\left( \sqrt {f} \right) {\it z_0}}
{b}}}}{\it z_0}
}
{
b{f}^{3/2}\sqrt {{\frac {1}{\sqrt {f}}}-1}\sqrt {{\frac {1}{\sqrt {f}}
}+1}
}
\quad  .
\label{probf}
\end{equation}
The average value  of  $f$  when
$b=83\,pc$ and
$z_0$=90  is  $\overline{f}=0.7927$.

\subsection{Monte Carlo Simulation}

\begin{figure*}
\begin{center}
\plotone{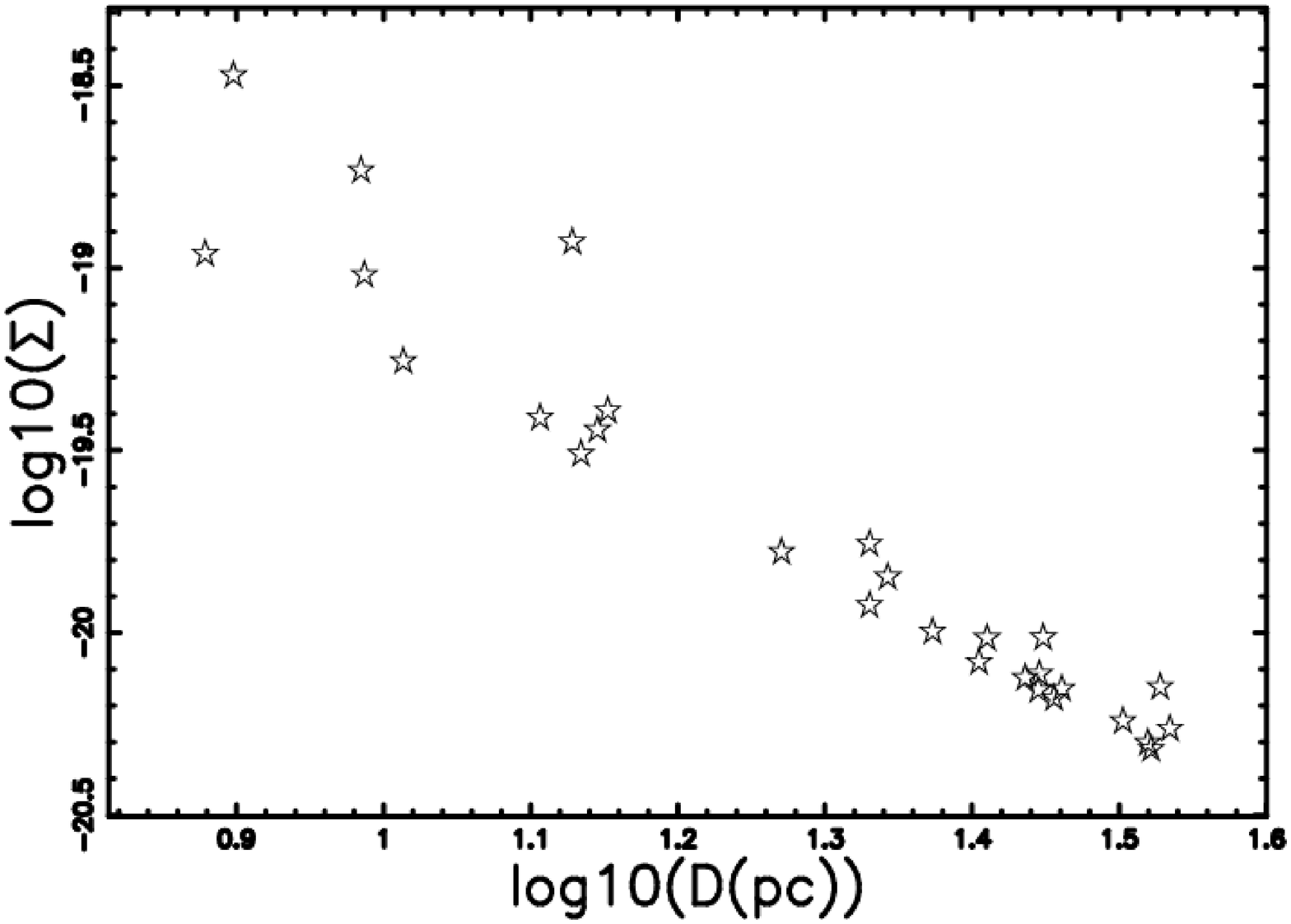}
\end {center}
\caption { A Monte Carlo   LOG-LOG \sigmad diagram at a frequency
of 1 GHz which simulates  the  Galactic SNRs for $D < 36.5 pc$.
The  parameters  are  $t_{max}=5055\, yr$  , $\alpha$       = 0.82
, $b$       = 83    $pc$ and $z_0$ = 90\, $pc$. The Monte Carlo
simulation gives $C_{\Sigma}$  = $2.71 10^{-17}$ and
$\beta_{\Sigma}$=2.45. }
          \label{mcteo}%
    \end{figure*}

\begin{figure*}
\begin{center}
\plotone{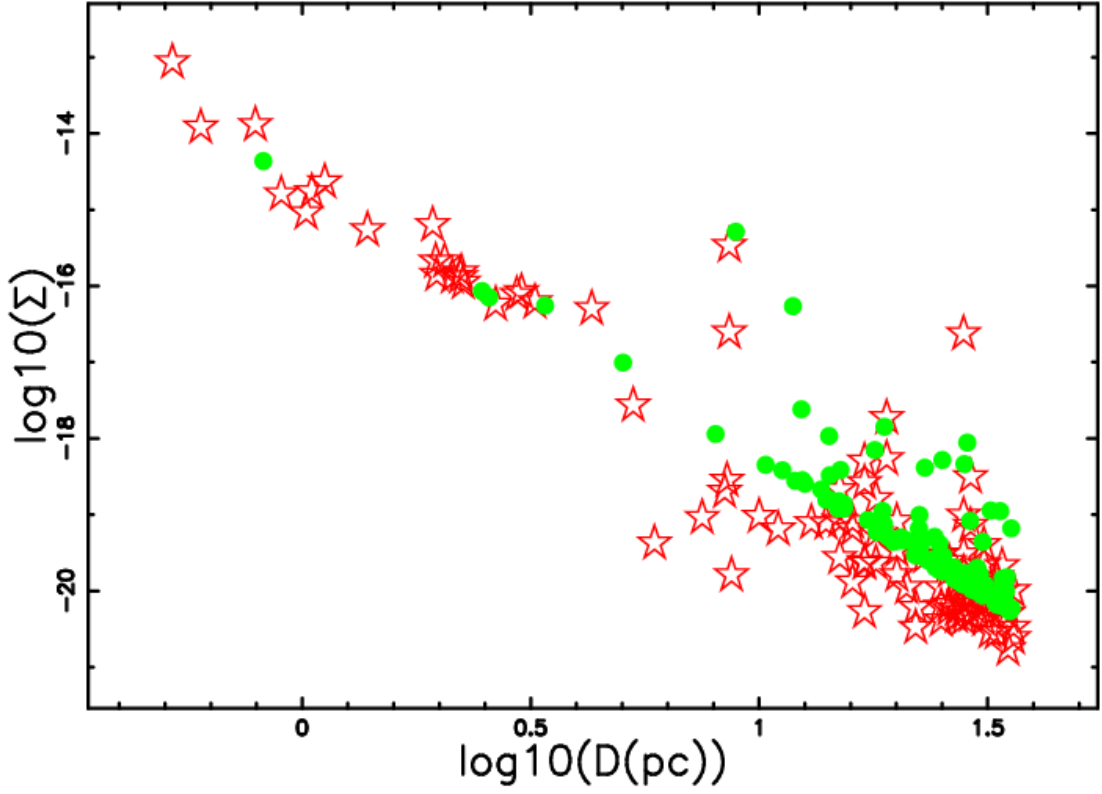}
\end {center}
\caption
{
A Monte Carlo   LOG-LOG \sigmad
diagram at a frequency of
1 GHz (green circles)
and
 extragalactic  SNRs  as given
by  \cite{Urosevic2005} for  $D < 36.5 pc$ (empty red stars). The
parameters  are  $t_{max}=2960 yr$  , $C=0.17$ , $\alpha$ =  0.57
, $b$       = 83    $\,pc$ and $z_0$ = 60 $\,pc$. The Monte Carlo
simulation gives $C_{\Sigma}$  = $8.84 10^{-16}$ and
$\beta_{\Sigma}$=3.65. }
          \label{mcextra}%
    \end{figure*}

We  are ready  to  build a
a Monte Carlo model for the
$\Sigma - D$ distribution of SNRs
for SNRs  characterized
by the  following constraints
\begin{itemize}
\item
The  lifetime  of a SNR  is
generated  between  0
and  $t_{max}$.
\item
The  time is converted in  diameter
, $D=R*2$ ,
adopting  Eq.  (\ref{radiustime}).
\item
A value of $f$ is randomly generated according to the PDF, Eq. (\ref{probf}).
\item
In the previous  points we have generated  one $D$
and  one $f$,     as a consequence    a value  of
$\Sigma_{1GHz}$   as  given by Eq.  (\ref{sigmadconf})
can be generated.
\end{itemize}
The  results  of this simulation  are displayed in Figure
\ref{mcteo} for the Galactic distribution, and in Figure
\ref{mcextra} for the extragalactic distribution. In this Section
we derived  the  profile  of the density in our galaxy as function
of the galactic height $z$  in the framework of the
self-gravitating disk. This new relationship allows us to find 
a new PDF that characterizes the effect of SNRs 
expanding into a medium of varying density.
The great fluctuations present in the
galactic \sigmad relationship are therefore explained by the SNR
evolution in a medium with a lower density.

\section{Conclusions}

The relation between radio surface  brightness and diameter of
galactic and extragalactic SNRs  presents a linear relationship in
the Log( $\Sigma$)-D plane. Superposed on  this observed linear
behavior there is a fluctuation
   which is due  to the variable density
with the galactic height  of the ISM. The major results  of this
paper are summarized as follows.
\begin{itemize}
\item The  power law  model  for the expansion of a SNR  coupled
with a decreasing  density  medium allows to build a simple
expression  for the luminosity as  well the radio surface
brightness, see  Eqs.  (\ref{surftheo}) and (\ref{lumtheo}). 
\item
A more sophisticated approach  to the law of motion as given by
the momentum conservation in a medium with variable density allows
to obtain the same  result  , see Eq. (\ref{sigmadconf}), once an
asymptotic behavior for the law  of motion is given, see Eq.
(\ref{asymptotic}). In this model  is possible to  introduce a
contrast parameter $f$  which allows  to state that both the
radius and  the velocity increase as $\frac{1}{f}$ and the surface
brightness as  $\frac{1}{f^3}$.
 \item The value of $f$  for each
SNR is given  by a new PDF (Eq. \ref{probf}) which is derived  from
the  observed  PDF for SNR as function of  the galactic height ,
see  Eq.  (\ref{pdfsnrz}) and from a physical profile  of density
of the ISM as  given by  Eq. (\ref{sech2}). \item A  Monte Carlo
simulation of the radio surface  brightness explains the
fluctuations visible  in the \sigmad relationship as a
probabilistic effect  to have a SNR with a given galactic height
or  $f$.
\end{itemize}

\section*{ Acknowledgements}
I would like to thank the anonymous referee for constructive
comments on the text.

\begin{thebibliography}{32}
\expandafter\ifx\csname natexlab\endcsname\relax\def\natexlab#1{#1}\fi

\bibitem[{{Bandiera} \& {Petruk}(2010)}]{Bandiera2010}
{Bandiera}, R., \& {Petruk}, O. 2010, \aap, 509, A34+

\bibitem[{{Bell}(1978{\natexlab{a}})}]{Bell_I}
{Bell}, A.~R. 1978{\natexlab{a}}, \mnras, 182, 147

\bibitem[{{Bell}(1978{\natexlab{b}})}]{Bell_II}
---. 1978{\natexlab{b}}, \mnras, 182, 443

\bibitem[{{Berezhko} \& {V{\"o}lk}(2004)}]{Berezhko2004}
{Berezhko}, E.~G., \& {V{\"o}lk}, H.~J. 2004, \aap, 427, 525

\bibitem[{{Bertin}(2000)}]{Bertin2000}
{Bertin}, G. 2000, {Dynamics of Galaxies} (Cambridge: {Cambridge University
  Press.})

\bibitem[{{Bisnovatyi-Kogan} \& {Silich}(1995)}]{Bisnovatyi1995}
{Bisnovatyi-Kogan}, G.~S., \& {Silich}, S.~A. 1995, \rmp, 67, 661

\bibitem[{{Clark} \& {Caswell}(1976)}]{Clark1976}
{Clark}, D.~H., \& {Caswell}, J.~L. 1976, \mnras, 174, 267

\bibitem[{{de Young}(2002)}]{deyoung}
{de Young}, D.~S. 2002, {The physics of extragalactic radio sources} (Chicago:
  University of Chicago Press)

\bibitem[{{Dickey} \& {Lockman}(1990)}]{Dickey1990}
{Dickey}, J.~M., \& {Lockman}, F.~J. 1990, \araa, 28, 215

\bibitem[{{Duric} \& {Seaquist}(1986)}]{Duric1986}
{Duric}, N., \& {Seaquist}, E.~R. 1986, \apj, 301, 308

\bibitem[{{Fermi}(1949)}]{Fermi49}
{Fermi}, E. 1949, Physical Review, 75, 1169

\bibitem[{{Fermi}(1954)}]{Fermi54}
---. 1954, \apj, 119, 1

\bibitem[{{Green}(2009)}]{Green2009}
{Green}, D.~A. 2009, Bulletin of the Astronomical Society of India, 37, 45

\bibitem[{{Gull}(1973)}]{Gull1973}
{Gull}, S.~F. 1973, \mnras, 161, 47

\bibitem[{{Guseinov} {et~al.}(2003){Guseinov}, {Ankay}, {Sezer}, \&
  {Tagieva}}]{Guseinov2003}
{Guseinov}, O.~H., {Ankay}, A., {Sezer}, A., \& {Tagieva}, S.~O. 2003,
  Astronomical and Astrophysical Transactions, 22, 273

\bibitem[{{Katsuda} {et~al.}(2010){Katsuda}, {Petre}, {Mori}, {Reynolds},
  {Long}, {Winkler}, \& {Tsunemi}}]{Katsuda2010}
{Katsuda}, S., {Petre}, R., {Mori}, K., {Reynolds}, S.~P., {Long}, K.~S.,
  {Winkler}, P.~F., \& {Tsunemi}, H. 2010, \apj, 723, 383

\bibitem[{{Kesteven}(1968)}]{Kesteven1968}
{Kesteven}, M.~J.~L. 1968, Australian Journal of Physics, 21, 739

\bibitem[{{Lang}(1999)}]{lang}
{Lang}, K.~R. 1999, {Astrophysical formulae. (Third Edition)} (New York:
  Springer)

\bibitem[{{Lequeux}(1962)}]{Lequeux1962}
{Lequeux}, J. 1962, Annales d'Astrophysique, 25, 221

\bibitem[{{Lockman}(1984)}]{Lockman1984}
{Lockman}, F.~J. 1984, \apj, 283, 90

\bibitem[{{Marcaide} {et~al.}(2009){Marcaide}, {Mart{\'{\i}}-Vidal}, {Alberdi},
  \& {P{\'e}rez-Torres}}]{Marcaide2009}
{Marcaide}, J.~M., {Mart{\'{\i}}-Vidal}, I., {Alberdi}, A., \&
  {P{\'e}rez-Torres}, M.~A. 2009, \aap, 505, 927

\bibitem[{{Mart{\'{\i}}-Vidal} {et~al.}(2011){Mart{\'{\i}}-Vidal}, {Marcaide},
  {Alberdi}, {Guirado}, {P{\'e}rez-Torres}, \& {Ros}}]{MartiVidalb2011}
{Mart{\'{\i}}-Vidal}, I., {Marcaide}, J.~M., {Alberdi}, A., {Guirado}, J.~C.,
  {P{\'e}rez-Torres}, M.~A., \& {Ros}, E. 2011, \aap, 526, A143+

\bibitem[{{McCray}(1987)}]{McCray1987}
{McCray}, R.~A. 1987, in Spectroscopy of Astrophysical Plasmas, ed.
  {A.~Dalgarno \& D.~Layzer} (Cambridge: {Cambridge University Press}),
  255--278

\bibitem[{{Milne}(1979)}]{Milne1979}
{Milne}, D.~K. 1979, Australian Journal of Physics, 32, 83

\bibitem[{{Padmanabhan}(2002)}]{Padmanabhan_III_2002}
{Padmanabhan}, P. 2002, {Theoretical astrophysics. Vol. III: Galaxies and
  Cosmology} ({Cambridge, MA}: {Cambridge University Press})

\bibitem[{{Poveda} \& {Woltjer}(1968)}]{Poveda1968}
{Poveda}, A., \& {Woltjer}, L. 1968, \aj, 73, 65

\bibitem[{{Press} {et~al.}(1992){Press}, {Teukolsky}, {Vetterling}, \&
  {Flannery}}]{press}
{Press}, W.~H., {Teukolsky}, S.~A., {Vetterling}, W.~T., \& {Flannery}, B.~P.
  1992, {Numerical Recipes in FORTRAN. The Art of Scientific Computing}
  (Cambridge: Cambridge University Press)

\bibitem[{{Shklovskii}(1960)}]{Shklovskii1960a}
{Shklovskii}, I.~S. 1960, \azh, 37, 256

\bibitem[{{Uro{\v s}evi{\'c}} {et~al.}(2005){Uro{\v s}evi{\'c}}, {Pannuti},
  {Duric}, \& {Theodorou}}]{Urosevic2005}
{Uro{\v s}evi{\'c}}, D., {Pannuti}, T.~G., {Duric}, N., \& {Theodorou}, A.
  2005, \aap, 435, 437

\bibitem[{{Winkler} {et~al.}(2003){Winkler}, {Gupta}, \& {Long}}]{Winkler2003}
{Winkler}, P.~F., {Gupta}, G., \& {Long}, K.~S. 2003, \apj, 585, 324

\bibitem[{{Xu} {et~al.}(2005){Xu}, {Zhang}, \& {Han}}]{Xu2005}
{Xu}, J., {Zhang}, X., \& {Han}, J. 2005, \cjaa, 5, 442

\bibitem[{{Zaninetti}(2011)}]{Zaninetti2011a}
{Zaninetti}, L. 2011, \apss, 333, 99

\end{thebibliography}

\newpage
\clearpage
\clearpage
\clearpage
\clearpage

\clearpage
\clearpage
\clearpage
\begin{table}
\caption
{
Numerical values of the parameters of the fits
for  \snr
 and
$\chi^2$. $N$  represents the number of  free parameters.
}
 \label{datafit}
 \[
 \begin{tabular}{ccc}
 \tableline
 \tableline
  N   & values  & $\chi^2$        \\
 \tableline
  &   power~law   &    \\
  2   & $\alpha$ = 0.82 $\pm$ 0.0048  & 6364 \\
 ~    &C = (0.015 $\pm$ 0.00011)    &  \\
~ &  nonlinear~radius  &   \\
  4   &
d=2.93; $r_{0}$ = 0.019~{pc}; &   276
 \\
   & $t_{0}$=0.249~{yr}; $v_0$ =100 000
$\frac{km}{s}$ &
\\
 \tableline
 \tableline
 \end{tabular}
 \]
 \end {table}
\clearpage
\begin{table}
\caption { Numerical values of the parameters of the fits
for  \sn1006
as  given by Eq. (\ref{twovar})
for  $R_{obs}$ = 9.59  pc ,
$V_{obs}$ =5441  $\frac{km}{s}$  and  t= 998 yr.
 }
 \label{datavar}
 \[
 \begin{tabular}{cc}
 \tableline
 \tableline
  unknown    & value        \\
 \tableline
 $\alpha$  & 0.579  \\

 C       & 0.1758     \\

 \tableline
 \tableline
 \end{tabular}
 \]
 \end {table}

\clearpage
\begin{table}
\caption { Numerical values of  $d$  }
 \label{datad}
 \[
 \begin{tabular}{cccc}
 \tableline
 \tableline
  environment  & $\beta_{obs}$ & $\alpha$=0.82    & $\alpha$=0.57       \\
 \tableline
extragalactic,~D  $<$  450 pc  & 3.1  & 2.44  & 0.83  \\
galactic ~~~~~~,~D  $<$  36.5 pc & 2.47 & 1.81  & 0.2  \\
 \tableline
 \tableline
 \end{tabular}
 \]
 \end {table}

\end{document}